# Relativistic potential energy of a bound particle

# (Non-dissipative relativistic harmonic oscillator)

J. Jahanpanah

Physics Faculty, Kharazmi University, 49 Mofateh Ave, 15719-14911, Tehran, Iran

jahanpanah@khu.ac.ir

**Abstract**

The well-known relation of Einstein's relativistic energy $E = mc^2$ for a free particle is extended to cover the total relativistic energy of a bound particle by calculating the relativistic potential energy. A non-dissipative harmonic oscillator (NDHO) is a fundamental bound system. Therefore, the potential energy of an NDHO is analytically extended from the nonrelativistic to the relativistic regime for the first time. This study is essentially concerned with the relativistic mass $m = \gamma m_0$, where the Lorentz factor $\gamma$ transforms the state of the second-order differential equation of an NDHO from linear into nonlinear. Although the nonlinear solution still remains periodic, the amplitude $A(\beta)$ and angular frequency $\omega(\beta)$ are determined only by the dimensionless factor $\beta = \dot{X}(0)/c$ (the ratio of the initial velocity $\dot{X}(0)$ to the light speed $c$). The time period $T(\beta) = 2\pi\omega^{-1}(\beta)$ asymptotically tends to infinity at the high initial relativistic velocities $\dot{X}(0) \to c$ (or $\beta \to 1$) in agreement with the principles of special relativity. It is demonstrated that the relativistic Hook force tends to zero when $\beta \to 1$ so that the state of the bound particle approaches that of a free particle. On the other hand, all relativistic relations are reduced to the corresponding nonrelativistic cases at low velocities $\beta \to 0$. The results are finally confirmed by demonstrating energy conservation since the sum of kinetic and potential energies remains constant throughout the nonrelativistic and relativistic regimes ($0 < \beta < 1$).

# 1. Introduction

The relativistic properties of a non-dissipative harmonic oscillator (NDHO) have been investigated using the principles of special relativity [1, 2]. From the beginning of the creation of relativity by Albert Einstein, he defined the relativistic kinetic energy of a free particle inspired by the equivalence of mass and energy as $E_k = (\gamma - 1)m_0 c^2$ [3, 4]. After a while, an important question was raised about the relation between the relativity and potential energy of a bound particle (system) [5, 6]. A harmonic oscillator is an ideal bound system because it is usually used to describe the physical behavior of more complicated systems such as Bose-condensate lithium atoms [7], hydrogen-like atoms [8], black-body radiation [9], and so on [10, 11]. To the best of my knowledge, nobody has ever introduced the relation between the relativistic restoring force and potential energy of a bound particle even in the simplest case that is the relativistic harmonic oscillator. Therefore, the core of the present research is to derive this relation.

Let us begin with the relativistic form of Newton's second law for an NDHO as $F_{rel} = dp_{rel}/d\tau = -kX$, where $k$ is the restitution coefficient of the force [12]. The relativistic linear momentum $p_{rel} = m\,dX/d\tau$ ($m = \gamma m_0$) plays the relativistic role through the Lorentz factor $\gamma = (1 - \dot{X}^2/c^2)^{-0.5}$. Using $\gamma$ and substituting $p_{rel}$ into $F_{rel}$, a second-order nonlinear differential equation is finally obtained for the motion of the relativistic NDHO as $\ddot{X} + \omega_0^2 \gamma^{-3} X = 0$, where $\omega_0 = \sqrt{k/m_0}$ is the natural frequency of a nonrelativistic harmonic oscillator [13, 14].

So far, many different numerical methods such as Harmonic balance [15, 16], differential transformation [17, 18], Adomian decomposition [19], and Homotopy perturbation [14, 20, 21] have been used to derive the approximate solutions for the second-order nonlinear differential equation of a relativistic NDHO. The aim of the present paper is to introduce an analytical solution for this equation without exploiting any numerical method for the first time. According to both numerical and analytical solutions, the periodic nature of harmonic oscillators is reserved even in the relativistic regime, while the physical quantities of amplitude $A(\beta)$ and angular frequency $\omega(\beta)$ are determined by the initial value of the relative velocity $\beta = \dot{X}(0)/c$.

The last aim of this study is to derive the relativistic form of the Hook force as well as the kinetic, potential, and mechanical energies of an NDHO. It is noteworthy that the relativistic Hook force is gradually transformed from the linear to the nonlinear state by increasing $\beta$ until it reaches zero at the ultimate relativistic limit $\beta = 1$. As a result, the alternation time $T(\beta)$ asymptotically goes to infinity so that the relativistic harmonic oscillator does not experience the initial velocities larger than the light speed ($\beta > 1$). Finally, the results are confirmed by demonstrating energy conservation. The sum of relativistic kinetic and potential energies remains constant and results in a straight line for the (total) mechanical energy throughout the possible values of $\beta$ ($0 < \beta < 1$).

## 2. Equation of motion and analytical solution

As explained in the preceding section, a relativistic NDHO obeys a second-order nonlinear differential equation of motion in the form [12, 14, 21]

$$\ddot{X} + \omega_0^2 \left(1 - \frac{\dot{X}^2}{c^2}\right)^{\frac{3}{2}} X = 0, \tag{1}$$

where X is the position of the oscillating particle; $\dot{X}$ and $\ddot{X}$ are the first and second derivatives of X with respect to time $\tau$. It is evident that Eq. (1) is nonlinear due to the Lorentz factor $\gamma^{-3} = (1 - \dot{X}^2/c^2)^{\frac{3}{2}}$ so that it immediately reduces to the well-known nonrelativistic linear equation $\ddot{X} + \omega_0^2 X = 0$ at classical velocities $\dot{X} \ll c$ ($\gamma = 1$). Now, if one defines the following dimensionless position and time quantities

$$x = \frac{\omega_0}{c} X \tag{2}$$

and

$$t = \omega_0 \tau \tag{3}$$

then Eq. (1) is simplified to

$$\ddot{x} + \left(1 - \dot{x}^2\right)^{\frac{3}{2}} x = 0. \tag{4}$$

Equation (4) has been approximated using many different numerical methods in the literature by considering the following initial position and velocity conditions [13-21]

$$x(0) = X(0) = 0 \tag{5}$$

and

$$\dot{x}(0) = \frac{\dot{X}(0)}{c} = \beta. \tag{6}$$

All the numerical methods agree on the periodic nature of Eq. (4), so we can define a trial solution that is consistent with the initial condition (5) as

$$x(t) = A(\beta) \sin\left[\frac{\omega(\beta)}{\omega_0} t\right], \tag{7}$$

where $A(\beta)$ and $\omega(\beta)$ are the amplitude and angular frequency of the relativistic NDHO, respectively. The amplitude $A(\beta)$ is easily obtained in terms of $\omega(\beta)$ by taking the derivative of trial solution (7) and using the initial condition (6) as

$$A(\beta) = \frac{\omega_0}{\omega(\beta)} \beta. \tag{8}$$

Now by substituting the first- and second-order derivatives of trial solution (7) into Eq. (4), a relation is obtained in terms of the time-independent variable $\omega(\beta)/\omega_0$ as

$$1 - \beta^2 \cos^2\left[\frac{\omega(\beta)}{\omega_0} t\right] = \left[\frac{\omega(\beta)}{\omega_0}\right]^{\frac{4}{3}}. \tag{9}$$

The static solution of Eq. (9) is given by

$$\frac{\omega(\beta)}{\omega_0} = (1 - \beta^2)^{\frac{3}{4}}, \tag{10}$$

where the only time-independent term of Maclaurin expansion $\cos^2\left[\frac{\omega(\beta)}{\omega_0} t\right]$ around $t = 0$ is rationally used to preserve the non-dissipative property of the relativistic harmonic oscillator. However, the accuracy of time-independent relation (10) is verified in section 5 by demonstrating energy conservation. In addition, the relation (10) is in complete agreement with the numerical solution (37) of Ref. [17], which has been derived by applying the inverse Laplace transformation to the Padé approximation [22].

The periodic behavior of the relativistic NDHO is illustrated in Fig. 1 for $\beta = 0.1$ (blue), 0.3 (red), 0.5 (green), and 0.7 (brown).

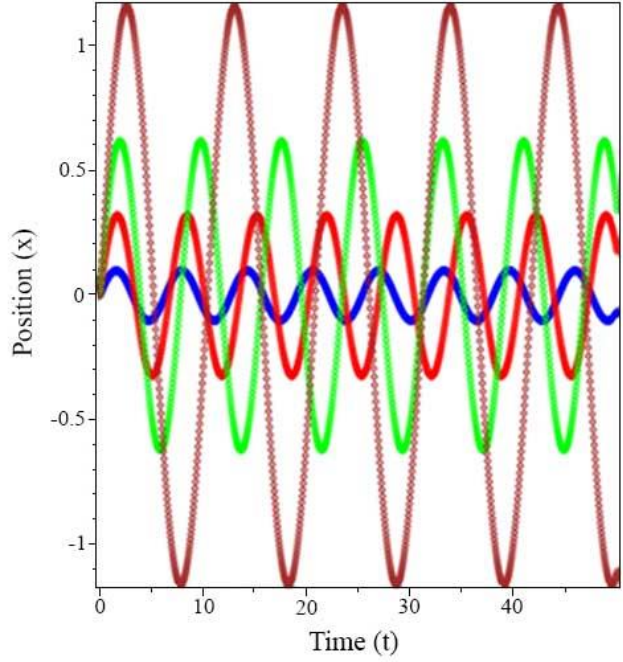

**Fig. 1.** The positional variations of the relativistic NDHO against the dimensionless time $t$ for $\beta = 0.1$ (blue), 0.3 (red), 0.5 (green), and 0.7 (brown).

It is evident that the amplitude and wavelength continuously increase by selecting higher values for the initial relative velocity $\beta = \dot{x}(0)$. Moreover, the normalized relativistic angular frequency $\omega(\beta)/\omega_0$ and the alternation time $T(\beta)/T_0$ tend to zero and infinity when $\beta$ approaches 1, as illustrated in Figs. 2(a) and 2(b), respectively. By increasing the initial relative velocity $\beta = \dot{x}(0)$, the relative alternation time $T(\beta)/T_0$ gradually increases in agreement with the time dilation principle of special relativity. Eventually, it diverges at $\beta = 1$ to avoid the breach of the upper velocity limit (light speed), i.e., for $\beta > 1$. The relativistic behavior of an NDHO is described in detail in the next section using the relativistic Hook force.

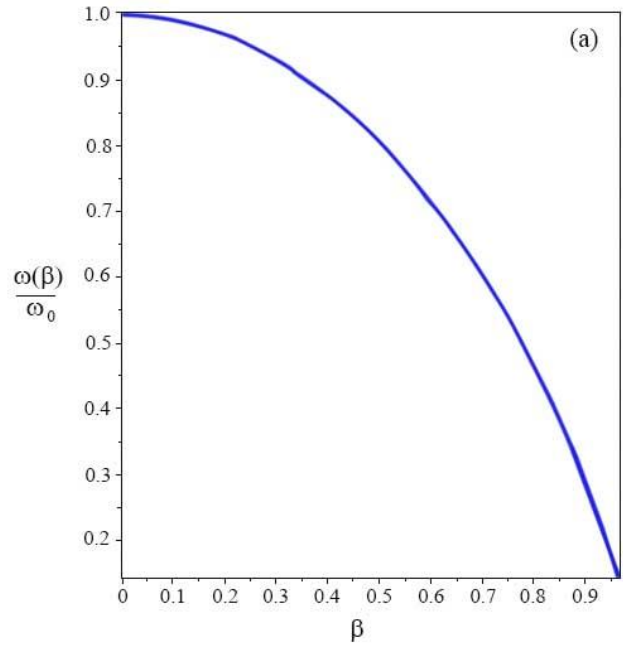

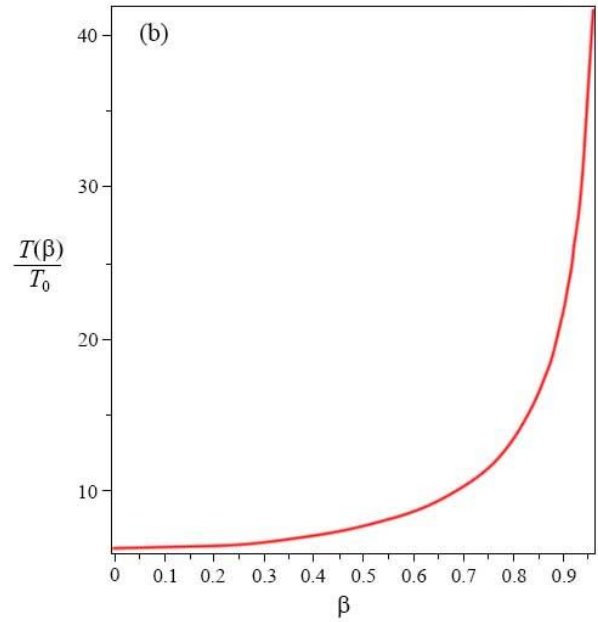

**Fig. 2.** The inverse variations of $\omega(\beta)/\omega_0$ and $T(\beta)/T_0$ are demonstrated in parts (a) and (b), respectively. These curves respectively tend to zero and infinity at the upper relativistic limit, $\beta = 1$, beyond which ($\beta > 1$) the light speed is exceeded.

## 3. Relativistic Hook force

The exclusive role of the Lorentz factor $\gamma = (1 - \dot{X}^2/c^2)^{-0.5}$ is to transform the second-order differential Eq. (4) from the nonrelativistic linear state ($\gamma = 1$) into the relativistic nonlinear state ($\gamma > 1$). Thus, this key factor automatically modifies the classical properties of physical quantities such as the linear momentum, the Hook force as well as the kinetic and potential energies. The first important quantity is the classical linear momentum $p = m_0 \dot{X}$, which is directly transformed to the corresponding relativistic relation by the Lorentz factor $\gamma$ as [1, 2]

$$p_{rel} = \gamma m_0 \dot{X} = m_0 c \dot{x}(1-\dot{x}^2)^{-0.5}, \tag{11}$$

where the relation (3) and derivative of relation (2) have simultaneously been used to derive the relation $\dot{X} = dX/d\tau = c\dot{x}$. The relativistic form of Newton's second law $F_{rel} = dp_{rel}/d\tau$ is now implemented to gain the relativistic force $F_{rel}$ in terms of the first- and second-order derivatives of dimensionless position $x$ as

$$F_{rel}(\dot{x}, \ddot{x}) = \frac{dp_{rel}}{d\tau} = \omega_0 \frac{dp_{rel}}{dt} = m_0 \omega_0 c (1-\dot{x}^2)^{-\frac{3}{2}} \ddot{x}. \tag{12}$$

In the next step, the general relation of relativistic force (12) is used for the special case of the relativistic NDHO. Consequently, the relativistic Hook force is obtained in terms of $x$ rather than $\dot{x}$ and $\ddot{x}$ by taking the first- and second-order derivatives of trial solution (7) and substituting the result into (12) as

$$F_{rel}(x) = -m_0 \omega_0 c \frac{\omega(\beta)^2}{\omega_0^2} x \left(1 - \beta^2 + \frac{\omega(\beta)^2}{\omega_0^2} x^2\right)^{-\frac{3}{2}}, \tag{13}$$

where the ratio of the relativistic to natural angular frequency $\omega(\beta)/\omega_0$ is given by the relation (10).

There are three important points regarding the relativistic nonlinear Hook force (13). First, many different numerical methods have been implemented to obtain an approximate analytical expression for $\omega(\beta)/\omega_0$ [13, 17, 21]. One can easily check the accuracy of these approximate methods by using the energy conservation relation presented in section 5. Second, the relativistic nonlinear Hook force (13) reduces to the nonrelativistic linear case at low initial velocities, i.e., $\beta \to 0$. In this case, $\omega(\beta)/\omega_0 \to 1$, so we have

$$F_{rel}(X) = \lim_{c \to \infty} \left[ -m_0 \omega_0^2 X \left(1 - \frac{\omega_0^2}{c^2} X^2 \right)^{-\frac{3}{2}} \right] = -m_0 \omega_0^2 X , \qquad (14)$$

where the relation (2) is used to replace the dimensionless position $x$ with the dimensional one X in (13). Third, the relativistic Hook force (13) approaches zero (the state of a free particle) at the initial velocities close to the light speed $\beta \to 1$, which leads to $\omega(\beta)/\omega_0 \to 0$ according to the relation (10).

The above three points are illustrated in Fig. 3 where the variations of the normalized relativistic Hook force $F_{rel}(x)/m_0 \omega_0 c$ are plotted versus the position $x$ for $\beta$ = 0.05 (blue), 0.365 (red), 0.68 (green), and 0.995 (brown).

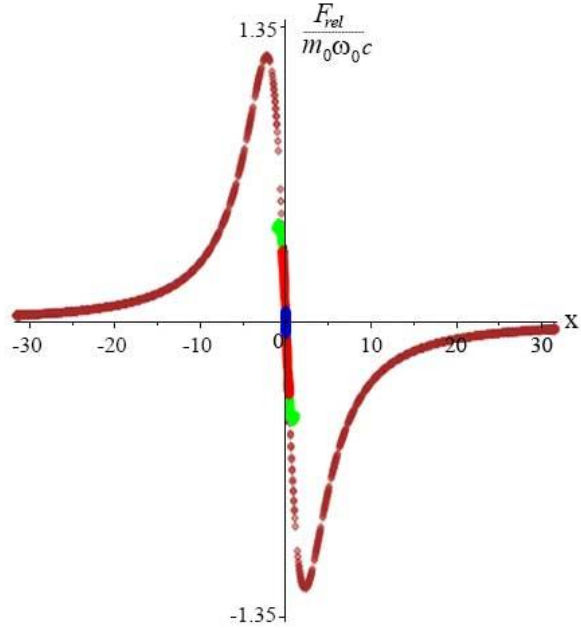

**Fig. 3.** The variations of the relativistic Hook force (13) versus the dimensionless position $x$ are shown for $\beta$ = 0.05 (blue), 0.365 (red), 0.68 (green), and 0.995 (brown). The relativistic Hook force gradually transforms from linear into nonlinear by increasing $\beta$.

The blue curve demonstrates the linear behavior of a relativistic NDHO at the classical initial velocities $\beta \ll 1$, while the red and green curves indicate the gradual deviation from the linear to the nonlinear state; until the nonlinear brown curve approaches zero at a high initial velocity, i.e., for $\beta \approx 0.995$. In the latter case ($\beta \to 1$), the relativistic NDHO seems to undergo state-change from the bound state to a free system (particle), but this cannot occur because the initial velocity $\dot{X}(0)$ never goes beyond the light speed $c$ ($\beta > 1$). Accordingly, the relative alternation time $T(\beta)/T_0$ tends to infinity, as illustrated in Fig. 2(b).

## 4. Relativistic kinetic and potential energies

The relativistic energy of a free particle has usually been defined based on the equivalence of mass and energy as [3, 23]

$$E_k = (\gamma - 1)m_0 c^2 = \left(p_{rel}^2 c^2 + m_0^2 c^4\right)^{\frac{1}{2}} - m_0 c^2. \tag{15}$$

If one now substitutes $p_{rel}$ from (11) to (15) and divides the result by the rest energy $m_0 c^2$, then an important relation is turned out for the dimensionless relativistic kinetic energy $T_{rel}$ as

$$T_{rel}(\dot{x}) = \left(1 - \dot{x}^2\right)^{-\frac{1}{2}} - 1, \tag{16}$$

which is indirectly dependent on the initial velocity $\beta = \dot{x}(0)$ through the derivative of the trial solution (7).

The relativistic potential energy $U_{rel}(x)$ is differently determined by using the well-known classical Newton's second law $F_{rel} = -\dfrac{\partial U_{rel}(x)}{\partial X} = -\dfrac{\omega_0}{c}\dfrac{\partial U_{rel}(x)}{\partial x}$ so that its integral gives an important novel relation (Jahanpana's relation) for the relativistic potential energy of a bound particle in the form

$$U_{rel}(x) - U_{rel}(0) = \int_0^x -\frac{c}{\omega_0} F_{rel}(\dot{x}, \ddot{x})\, dx, \tag{17}$$

where the relativistic force $F_{rel}(\dot{x}, \ddot{x})$ must be separately calculated by the relation (12) as a function of position $x$.

For example, consider the relativistic Hook force (13) which has been determined in terms of variable $x$ by substituting the first- and second-order derivatives of the trial solution (7) into the relativistic force relation (12). Consequently, the normalized relativistic potential energy $u_{rel}(x)$ is obtained for the relativistic NDHO by substituting (13) into (17) as

$$u_{rel}(x) = \frac{U_{rel}(x)}{m_0 c^2} = (1-\beta^2)^{\frac{-1}{2}} - \left(1 - \beta^2 + \frac{\omega(\beta)^2}{\omega_0^2} x^2 \right)^{\frac{-1}{2}}, \qquad (18)$$

where $x=0$ is the coordinate origin for $u_{rel}(x)$. It is easy to show that the relativistic potential energy (18) is reduced to the dimensionless classical potential energy $u_{rel}(x) = 0.5 x^2$ at low initial velocities (for $\beta \to 0$, that is, when $\omega(\beta)/\omega_0 \to 1$) based on relation (10). It should be noticed that the expansion $(1+x^2)^{-1/2} \approx 1 - 0.5 x^2$ is applicable because the ratio $\frac{x^4}{x^2} = \frac{\omega_0^2}{c^2} \frac{X^4}{X^2}$ tends to zero when the light speed $c$ tends to infinity. In addition, $u_{rel}(x)$ approaches zero (potential of a free particle) as $\beta$ approaches 1, which is consistent with the relativistic force (13).

## 5. Energy conservation as an evaluation tool for approximate methods

In the absence of any dissipation (loss), the Hook force retains its conservative property over the entire range of nonrelativistic and relativistic initial velocities $0 < \beta < 1$. The relativistic (total) mechanical energy $E$, which is the sum of relativistic kinetic $T_{rel}(\dot{x})$ and potential $u_{rel}(x)$ energies, thus obeys the Hook force, so we have the energy conservation relation in the form

$$T_{rel}(\dot{x}) + u_{rel}(x) = E = cte. \qquad (19)$$

The relativistic mechanical energy $E$ can now be calculated by substituting $T_{rel}(\dot{x})$ from (16) and $u_{rel}(x)$ from (18) into the energy conservation (19), and applying the initial conditions (5) and (6) as

$$E = (1 - \beta^2)^{-\frac{1}{2}} - 1. \qquad (20)$$

It is evident that the initial relative velocity $\beta = \dot{X}(0)/c = \dot{x}(0)$ can directly determine the relativistic mechanical energy of an NDHO before the oscillation onset. The relativistic mechanical energy $E$ approaches zero at the classical limit $\beta \to 0$, while it tends to infinity for $\beta \to 1$ based on the principle of equivalence of mass and energy ($\gamma \to \infty$). The variations of dimensionless relativistic potential (blue), kinetic (red), and mechanical (green) energies against $t$ and $x$ are illustrated respectively in Figs. 4(a) and 4(b) for $\beta = 0.8$. The sum of relativistic potential and kinetic energies against $t$, in a periodic shape, and against $x$, in a parabolic shape, gives a straight line, which represents the relativistic mechanical energy with a value of 0.667. This result is in complete agreement with the relation (20).

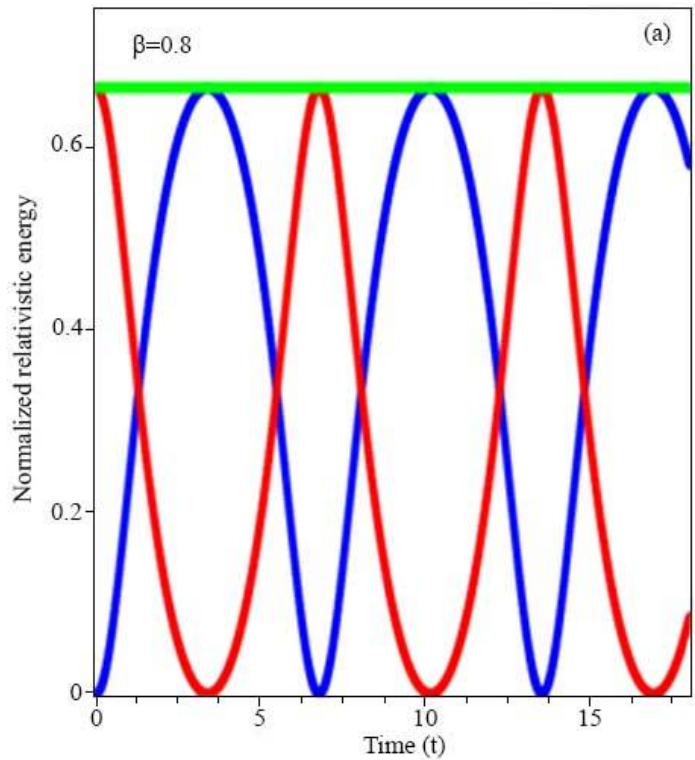

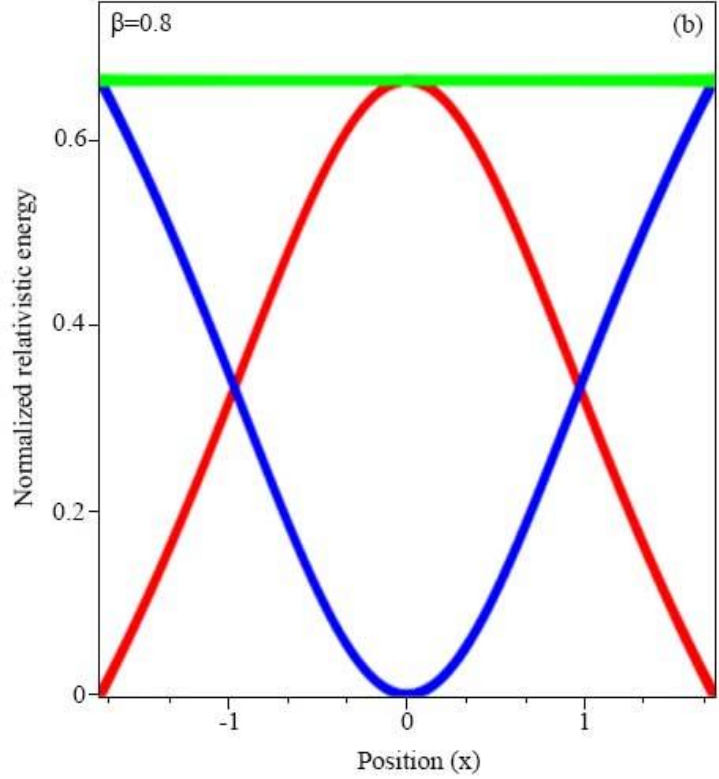

**Fig. 4.** The variations of relativistic kinetic (red) and potential (blue) energies of an NDHO against (a) dimensionless time with a periodic structure and (b) dimensionless position with a parabolic structure for $\beta = 0.8$. The sum of relativistic kinetic and potential energies gives a straight line (green) representing the relativistic (total) mechanical energy.

Two important points should be noted. First, the relativistic potential energies (17) and (18) are confirmed here based on the energy conservation relation (19). Second, the exceptional accuracy of the analytical trial solution (7) is confirmed by the straight line representing the relativistic mechanical energy in Figs. 4(a) and 4(b). As a result, the accuracy of the numerical methods can be evaluated by the energy conservation relation (19). For example, the second-order nonlinear differential equation (4) has also been solved by Mickens [13, 17] using the harmonic balance method (HBM) as

$$x(t) = \frac{\beta}{\omega}\left[1 + \frac{\beta^2}{8} + \frac{3\beta^4}{64}\right]\sin(\omega t) - \frac{\beta^3}{24\omega}\left[1 + \frac{3\beta^2}{128}\right]\sin(3\omega t) + \left[\frac{3\beta^5}{640\omega}\right]\sin(5\omega t), \qquad (21)$$

with $\omega$ given as

$$\omega = \frac{\omega(\beta)}{\omega_0} = \left(\frac{2-2\beta^2}{2-\beta^2}\right)^{\frac{1}{4}}. \tag{22}$$

The temporal variations of relativistic potential, kinetic, and mechanical energies are again plotted in Fig. 5 for the approximate solution (21) and the typical value $\beta = 0.8$. It is noteworthy that HBM gives a straight line for $E$ in the range $0 < \beta \leq 0.1$, but it gradually deviates from the straight line and shows a periodic behavior for larger values of $\beta$ in the range $0.1 < \beta < 1$ (see Fig. 5 plotted for $\beta = 0.8$).

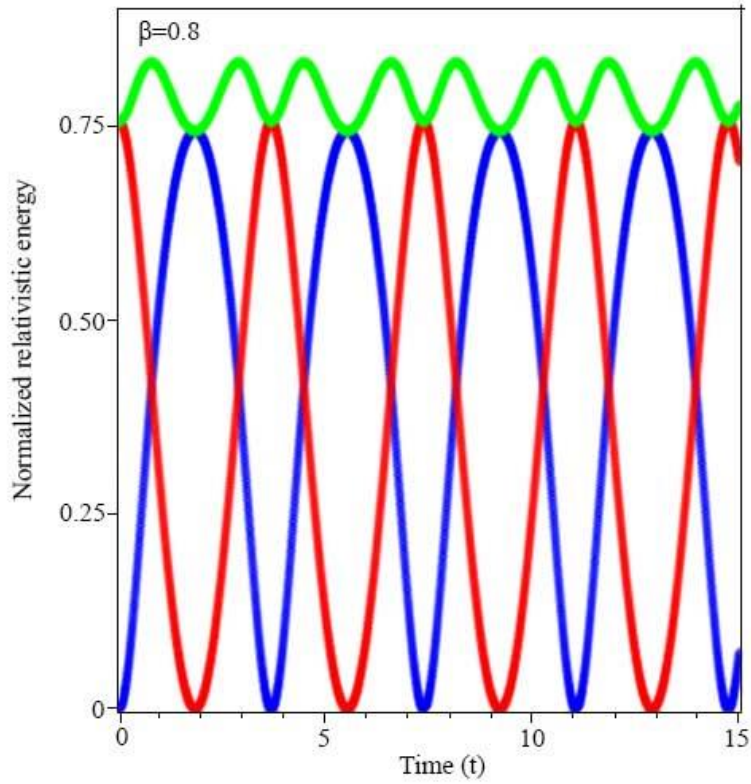

**Fig. 5.** The accuracy of Mickens's solution (21) is demonstrated for the typical initial velocity $\beta = 0.8$. The mechanical energy gradually deviates from a straight line for $\beta > 0.1$.

## 6. Conclusion

After announcing $E = mc^2$ as the total relativistic energy of a free particle, a new mystery was posed about the total relativistic energy of a bound particle. Since the relativistic kinetic energy of both free and bound particles is defined by $E_k = (\gamma - 1)m_0 c^2$, this mystery is all about the relativistic potential energy $U_{rel}(x)$ of a bound particle and its relevance to the relativistic force $F_{rel}(x)$. The main aim of the present research is to solve this mystery.

The only clue was the classical relation $F(x) = -\partial U(x)/\partial x$, which could be hold in the relativity with a similar structure to $F_{rel} = -\partial U_{rel}(x)/\partial x$, provided that $F_{rel}$ is substituted by the relativistic Newton's second law $F_{rel} = dp_{rel}/d\tau$. By taking the integral of $F_{rel} = -\partial U_{rel}(x)/\partial x$, the heuristic relation (17) is derived as the relativistic potential energy of a bound particle. The non-dissipative harmonic oscillator has then been considered as a fundamental bound system. By applying the trial solution (7) to the second-order nonlinear differential equation (4), the relativistic Hook force (13) and the potential energy (18) were obtained in terms of the initial relative velocity $\beta = \dot{X}(0)/c = \dot{x}(0)$. As a result, the numerical value of $\beta$ determines the physical behavior of a relativistic NDHO just before oscillation onset.

The classical relations can be restored by tending $\beta$ toward zero ($c \to \infty$). In contrast, at the upper limit of the relativistic range where $\beta \to 1$, the relative alternation time $T(\beta)/T_0$ suddenly diverges to avoid the impossible case of $\beta > 1$. At the same time, the relativistic Hook force approaches the state of a free particle by tending toward zero. The accuracy of the analytical and numerical solutions was eventually evaluated based on the energy conservation principle.

**References**

[1] Miller, A. I. *Albert Einstein's Special Theory of Relativity: Emergence (1905) and Early Interpretation (1905-1911),* 1998th ed, Springer-Verlag, 1997.

[2] McMahon, D. and Alsing, P. M. *Relativity Demystified - A self-teaching guide,* 1st ed, The McGraw-Hill Companies, 2005.

[3] Eugene, H. How Einstein confirmed $E_0 = mc^2$. Am. J. Phys. **79**, 591-600 (2011).; Eugene, H. How Einstein discovered $E_0 = mc^2$. The Physics Teacher, **50**, 91-94 (2012).

[4] Hecht, E. Einstein on mass and energy. Am. J. Phys. **77**, 799-806 (2009).

[5] L'e'nigme, L. $E = Mc^2$: Energie potentielle et renormalization de la masse. J. Phys. Radium. **25**, 883 (1964).

[6] Brillouin, L. The actual mass of potential energy: A correction to classical relativity. Proceedings of the National Academy of Sciences. **53** (3), 475-482 (1965).

[7] Fujiwara, K. M. et al. Experimental realization of a relativistic harmonic oscillator. New J. Phys. **20**, 063027 (2018).

[8] Jahanpanah, J. The forming mechanism of spontaneous emission noise flux radiated from hydrogen-like atoms by means of vibrational Hamiltonian. AIP Advances. **11**, 035203 (2021).

[9] Friedman, Y. Digitization of the harmonics oscillator in extended relativity. Phys. Scr. **87**, 065702 (2013).

[10] Big-Alabo, A. Continuous piecewise linearization method for approximate periodic solution of the relativistic oscillator. IJMEE. **48**, 178-194 (2020).

[11] Tung, M. M. The relativistic harmonic oscillator in a uniform gravitational field. Mathematics. **9**, 294 (2021).

[12] Hutten, E. H. Relativistic (Non-Linear) oscillator. Nature. **205**, 892 (1965).

[13] Mickens, R. E. Periodic solutions of the relativistic harmonic oscillator. JSV. **212**(5), 90-908 (1998).

[14] Biazar, J. and Hosami, M. An easy trick to a periodic solution of relativistic harmonic oscillator. JOEMS. **22**, 45-49 (2014).

[15] Mickens, R. E. *Oscillations in planar dynamic systems*, Word Scientific, Singapore, 1996.

[16] Belendez, A. Pascual, C. Mendez, D. I. Neipp, C. Solutions of the relativistic (an) harmonic oscillator using the harmonic balance method. JSV. **311**, 1147-1456 (2008).

[17] El-Halim Ebaid, A. Approximate periodic solutions for the non-linear relativistic harmonic oscillator via differential transformation method. Communication in Nonlinear Science and Numerical Simulation. **15**, 1921-1927 (2010).

[18] Momani, S. and Suat Erturk, V. Solutions of non-linear oscillators by the modified differential transform method. Comput. Math. With Appl. **55**, 833-842 (2008).

[19] Gonzalez-Gaxiola, O. Santiago, J. A. and Ruiz de Chavez, J. Solutions for the nonlinear relativistic oscillator via Laplace-Adomian decomposition method. In. J. Appl. Comput. Math. **3**, 2627-2638 (2017).

[20] Cveticanin, L. Homotopy-perturbation method for pure nonlinear differential equation. Chaos, Solitons and fractals. **30**, 1221-1230 (2006).

[21] Belendez, A. and et al. Higher-order approximation solutions to the relativistic and duffing-harmonic oscillators by modified He's homotopy methods. Phys.scr. **77**, 025004 (2008).

[22] Wu, S. and Bercu, G. Pade approximation for inverse trigonometric functions and their applications. JIA. **31**, 1-12 (2017).

[23] Mundarain, D. About the non-relativistic limit of the phase velocity of matter waves. Eur. J. Phys. **38**, 045402 (2017).